\documentclass[12pt,thmsa]{article}

\input{tcilatex}

\begin{document}

\author{A. Yu. Zyuzin \\
A.F.Ioffe Physical-Technical Institute 194021 Saint-Petersburg, Russia}
\title{Superfluorescent decay in quasiballistic disordered systems.}
\maketitle

\begin{abstract}
Superfluorescent decay in a weakly disordered slab is considered. By solving
the Maxwell-Bloch equations we show, that the cooperation number increases
with increasing of the mean free path. Superfluorescent impulse is emitted
at a small angle to the slab, isotropically in a plane direction.

PACS.42.20-Propagation and transmission in inhomogeneous media.

PACS.42.20G-Scattering, diffraction and polarization.
\end{abstract}

\textbf{Introduction.}

During the spontaneous emission of uncorrelated exited atoms, their dipole
moments interact each other via the emitted electromagnetic field. If the
lifetime of exited state is long enough, the radiation can induce the
correlation between the group of large number $N_c$ (so called cooperation
number) atoms. As a result of correlation, $N_c$ exited atoms emit as one
macroscopic dipole. Thus, the decay of the uncorrelated exited atoms, which
starts with the low level of emitted radiation during the developing of
macroscopic dipole, is continued by the emission of this macroscopic dipole.
At this last stage the duration of spontaneous decay decreases by the factor 
$N_c^{-1}$ and the maximum of intensity of emitted impulse increases by the
factor $N_c^2$ compare that parameters of spontaneous decay of a single atom 
\cite{Dicke54} . The total process is called superfluorescence \cite
{Bonifatio75} .

In extended system the cooperation number depends on the geometry of the
sample. For a example, in a well studied pencil-like sample it is $N_c\sim
\rho \lambda ^2L$ in one mode approximation \cite{Bonifatio75}, \cite
{Arecchi70} . Here $\rho $ is density of the active atoms. $L$ is the length
of the sample, which diameter is much larger than the wavelength of
radiation $\lambda .$ Maximum of impulse intensity is $\sim N_c^2\times
\frac N{N_c}$, where $N$ is total number of the exited atoms. The factor $%
\frac N{N_c}\gg 1$ can be interpreted as a number of macroscopic dipoles in
the sample.

Correlation between the exited atoms in this case develops due to the free
propagation of the electromagnetic field along the pencil-like sample.
During the propagation, the field can be absorbed and re-emitted. These
processes put an upper limit on the geometrical size $L,$ and therefore on $%
N_c$ \cite{Bonifatio75}, \cite{Arecchi70} .

The case, when the superfluorescence occurs in a disordered system, was
considered in \cite{zyuzin98} . It was shown that in the disordered slab of
the thickness $L\gg l$ ( $l$ is the mean free path of the radiation due to
scattering by the static fluctuations of dielectric constant) the
cooperation number is $N_c\sim \rho \lambda ^2\frac{L^2}l$. Here factor $%
\frac{L^2}l$ is the length of diffusive trajectory of the radiation.
Scattering prevents the escape of the radiation from a system with the
dimensions larger than the mean free path, therefore the cooperation number
increases with the increasing of disorder up to the upper limit where
absorption and re-emission starts. The maximum $N_c$ is of order of that in
a pure case.

The cooperation number is a random quantity in the disordered system.
Approximations adopted in \cite{zyuzin98} in order to obtain the analytical
solution allows calculation only of the average value of the cooperation
number.

Here we consider the superfluorescence of the weakly disordered slab with
the thickness $L$ less than the radiation mean free path. We show that the
cooperation number in this system is $N_c\sim \rho l\lambda ^2$. The
correlation develops due to the propagation of the field along slab.
Fluctuations of the dielectric constant scatter radiation out of volume,
which contains the active atoms, thus decreasing the cooperation number.
Superfluorescent impulse is emitted at small angle to the slab and
isotropically in a plane direction.

\textbf{Maxwell-Bloch equations.}

We consider scalar version of problem. The coupling between the polarization
density $\frac 12\left\{ e^{i\omega t}P\left( \vec{r};t\right) +e^{-i\omega
t}P^{*}\left( \vec{r};t\right) \right\} $, the population difference density 
$\Delta N\left( \vec{r};t\right) $, and the field $\frac 12\left\{
e^{i\omega t}E\left( \vec{r};t\right) +e^{-i\omega t}E^{*}\left( \vec{r}%
;t\right) \right\} $ can be described by the classical Maxwell-Bloch
equations. In this approach amplified spontaneous emission noise is
neglected, which is a good approximation for the superfluorescence \cite
{Bonifatio71}. $P\left( \vec{r};t\right) $ and $E\left( \vec{r};t\right) $
are slowly time-varying complex quantities. $\omega =ck$ is the atomic
frequency.

First two Maxwell-Bloch equations have the form~\cite{Siegman86}

\begin{equation}
\frac d{dt}P\left( \vec{r};t\right) =\frac{i\left| \mu \right| ^2}\hbar
\Delta N\left( \vec{r};t\right) E\left( \vec{r};t\right)  \label{eq1}
\end{equation}

\begin{equation}
\frac d{dt}\Delta N\left( \vec{r};t\right) =-\frac i{2\hbar }\left\{
P^{*}\left( \vec{r};t\right) E\left( \vec{r};t\right) -P\left( \vec{r}%
;t\right) E^{*}\left( \vec{r};t\right) \right\} .  \label{eq2}
\end{equation}

Here $\mu $ is the electric dipole moment.

\strut It is assumed that the population inversion relaxation time and the
dephasing time are larger than the delay time of the superfluorescent
impulse $t_0$, which will be defined lately in the equation (9). We also
neglect inhomogeneous broadening.

The population difference $\Delta N\left( \vec{r};t\right) $ and the
polarization $\frac{P\left( \vec{r};t\right) }\mu $ in the equations (\ref
{eq1}) and (\ref{eq2}) are the components of the local Bloch vector averaged
over scales smaller than the wavelength of the radiation $\lambda $. We
choose the initial conditions $\overline{P\left( \vec{r},t=0\right) }=0$ and 
$\overline{P\left( \vec{r},t=0\right) P^{*}\left( \vec{r}^{\prime
},t=0\right) }=\rho \left| \mu \right| ^2\delta \left( \vec{r}-\vec{r}%
^{\prime }\right) $, which describe the initially uncorrelated atoms. The
line means averaging over the initial state. $\vec{r}=\left( \vec{\rho}%
,z\right) $ , $\vec{\rho}$ is in-plane coordinate. The population difference
and the polarization are confined in a region of the slab $\left| z\right|
\leq \frac L2$.

The field wave equation for the slow time-varying component $E\left( \vec{r}%
;t\right) $ has the form

\begin{equation}
i\epsilon \left( \vec{r}\right) \frac d{dt}E\left( \vec{r};t\right) -\left\{
-\frac{c^2}{2\omega }\Delta -\frac{\omega \epsilon \left( \vec{r}\right) }%
2\right\} E\left( \vec{r};t\right) =2\pi \omega P\left( \vec{r};t\right) .
\label{eq3}
\end{equation}

Here $\epsilon \left( \vec{r}\right) =1+\delta \epsilon \left( \vec{r}%
\right) $ is the dielectric function of the medium which contains active
atoms. $\delta \epsilon \left( \vec{r}\right) $ causes scattering.

\textbf{Calculation of cooperation number.}

We consider scattering by the small fraction of the particles of the size
less than $\lambda $ embedded into the active medium. This is experimentally
relevant case and for the theoretical treatment this case has an advantage
that $l$ is equal to the transport mean free path.

Averaged over random positions of the scattering particles equation for the
field has the form of the equation (\ref{eq3}) with substitution of dumping $%
\frac i{kl}$ for $\delta \epsilon \left( \vec{r}\right) $ \cite{Abrikosov63}.

To consider superfluorescence we neglect time derivative in equation for the
field at the very beginning. This usual approximation means that the time of
the escape of radiation $\frac lc$ from the system is smaller, than the time
of exchange energy between field and atoms.

Solving equation (\ref{eq1}) we obtain equation for the field, averaged over
the disorder $\left\langle E\left( \vec{r};t\right) \right\rangle $

\begin{equation}
\left\{ \Delta +k^2+\frac{ik}l\right\} \left\langle E\left( \vec{r};t\right)
\right\rangle =4\pi k^2P\left( \vec{r};t=0\right) +\frac{4\pi ik\left| \mu
\right| ^2}\hbar \int\limits_0^tdt\Delta N\left( \vec{r},t\right)
\left\langle E\left( \vec{r};t\right) \right\rangle  \label{eq4}
\end{equation}

To find population difference from equation (2) we need to solve equation
(4) inside the slab. Below we consider the case when $l>kL^2$. This allows
to neglect dependence of $\Delta N\left( \vec{r},t\right) $ on coordinate $%
\vec{r}$. We assume that at initial moment $\Delta N\left( \vec{r}%
,t=0\right) =\rho $ , where $\rho $ is density of the active atoms. The
quantity $\left\langle E\left( \vec{r};t\right) \right\rangle \left\langle
P^{*}\left( \vec{r};t\right) \right\rangle $ weakly depends on coordinate
when mean free path is larger than $l>kL^2$. This allows to neglect the
dependence of $\Delta N\left( \vec{r},t\right) $ during the decay.

From equation (\ref{eq4}) it follows that the time dependence of field is
determined by the quantity 
\begin{equation}
\chi \equiv \frac 2{\rho \omega \tau _0^2}\int\limits_0^tdt\Delta N\left(
t\right)  \label{eq5}
\end{equation}

where $\tau _0\equiv \sqrt{\frac \hbar {2\pi \rho \omega \left| \mu \right|
^2}}=\sqrt{\frac{4\pi \tau _{rad}}{3\omega \rho \lambda ^3}}$ is the
characteristic time of energy exchange between the field and the atomic
system~\cite{Bonifatio75}$.$ $\tau _{rad}^{-1}\equiv \frac{8\pi ^2\left| \mu
\right| ^2}{3\hbar \lambda ^3}$ is the radiative decay time of a single atom.

Performing Laplace transformation of equation (\ref{eq4}) over $\chi $ and
Fourier transformation over in plane coordinate $\vec{\rho}$ we obtain
solution inside the slab

\begin{equation}
\left\langle E\left( \vec{q};s\right) \right\rangle =\frac{2\pi k^2}{%
is\left[ p+\frac{kL}{2l}\left( 1-\frac{kl}s\right) \right] }%
\int\limits_{-\frac L2}^{\frac L2}dz^{\prime }P\left( \vec{q},z^{\prime
};t=0\right) ,  \label{eq6}
\end{equation}

where $\left\langle E\left( \vec{q};s\right) \right\rangle =\int d^2\vec{\rho%
}exp\left( i\vec{q}\vec{\rho}\right) \int\limits_0^\infty d\chi exp\left(
-s\chi \right) \left\langle E\left( \vec{r};\chi \left( t\right) \right)
\right\rangle $,

$p=\sqrt{k^2-q^2}$ , $Im\left( p\right) >0$ .

The inverse Laplace transformation of (\ref{eq6}) gives an expression

\begin{equation}
\left\langle E\left( \vec{q};\chi \right) \right\rangle =\frac{2\pi k^2}{%
i\left[ p+\frac{kL}{2l}\right] }exp\left[ \frac{Lk^2\chi }{2\left( p+\frac{kL%
}{2l}\right) }\right] \int\limits_{-\frac L2}^{\frac L2}dz^{\prime
}P\left( \vec{q},z^{\prime };t=0\right) .  \label{eq7}
\end{equation}

Performing the inverse Fourier transformation and taking into account the
initial conditions for the polarization density, we obtain

\begin{quotation}
\begin{equation}
\frac{\overline{\left\langle E\left( \vec{r};\chi \right) \right\rangle
\left\langle P^{*}\left( \vec{r};\chi \right) \right\rangle }}{i\rho \left|
\mu \right| ^2k^3}\simeq -\frac L{8kl^2\chi }exp\left( 2kl\chi \right) .
\label{eq8}
\end{equation}
\end{quotation}

Expression (\ref{eq8}) gives asymptotic for $kl\chi >1$.

Equation (\ref{eq2}) and expression (\ref{eq8}) allow to solve for the
population difference. Neglecting time-dependence of slow varying pre-factor
in expression (8) we obtain for the population difference 
\begin{equation}
\Delta N\left( \vec{r},t\right) =\rho \tanh \frac{t_0-t}{2\tau _N},
\label{eq9}
\end{equation}

where $\tau _N\equiv \frac{\tau _{rad}}{N_C}$ and cooperation number is equal

\begin{equation}
N_C=6\rho \lambda ^2l.  \label{eq10}
\end{equation}

It is of order of number of active atoms in a tube with cross section $%
\lambda ^2$ and length $l$.

The delay time $t_0\simeq \tau _Nln\frac{\rho \lambda ^2l^2}L>>\tau _N$ is
calculated by matching (\ref{eq8}) and (\ref{eq9}) with initial case when $%
\chi =\frac{2t}{\omega \tau _0^2}$ and assuming that at the beginning of
collective decay ( $kl\chi \sim 1$) population difference does not change.

Let us note that expression (\ref{eq9}) coincides with the standard
expression for the population difference in Markovian theory of
superfluorescence.

To obtain the expression for emitted radiation we must solve equation (\ref
{eq4}) for $\left| z\right| >\frac L2$ and given time-dependent population
difference (\ref{eq9}).

Intensity of radiation from a unite volume of slab can be represented as 
\begin{equation}
I\left( \vec{r};t\right) =I_0\left( \vec{r}\right) \left[ \frac{2l}L\frac{%
\sin \theta }{1+\frac{2l}L\sin \theta }\right] ^2\left[ \frac{\cosh \frac{t_0%
}{2\tau _N}}{\cosh \frac{t-t_0}{2\tau _N}}\right] ^{\frac 2{1+\frac{2l}L\sin
\theta }}  \label{eq11}
\end{equation}

Here $I_0\left( \vec{r}\right) $ is the intensity of the ordinary
fluorescence of a unite volume with exited atoms density $\rho $ at distance 
$\vec{r}$. It is isotropic function $I_0\left( \vec{r}\right) \sim r^{-2}$.

The second factor in (\ref{eq11}) accounts for the mean free path of
radiation in the medium, $\sin \theta =\frac{\left| z\right| }r$.

The last factor in (\ref{eq11}) describes the angular and time dependences
of superfluorescent impulse.

\textbf{Conclusion.}

We calculated the cooperation number and the intensity of cooperative decay
in weakly disordered slab. Expressions (10) and (11) are the main results of
the paper.

It is shown that volume of the cooperating regions is proportional to volume
of a tube with length, which is equal mean free path, and of cross-section
dimension of order of the wavelength. The maximum of radiation is emitted in
a small angle to the surface of slab isotropically in a plane direction.

While calculation of (10) and(11) was performed under the condition $l>>kL^2$
it is reasonable to assume, that the scattering limits the cooperation
number until $l\simeq L$.

\textbf{Acknowledgment.}

This work was supported by the Russian Fund for Fundamental Research under
Grant number 97-02-18078.

\end{document}